\documentclass[10pt, a4paper]{article}
\usepackage{amsmath}
\usepackage{amsfonts}
\usepackage{fullpage}
\usepackage{hyperref}
\usepackage[dvipsnames]{xcolor}
\usepackage{graphicx}
\usepackage{array}
\usepackage{makecell}
\usepackage{tabularx}
\usepackage{multirow}
\usepackage{nicematrix}
\usepackage{soul}
\usepackage{enumitem,calc}
\usepackage{tcolorbox}
\usepackage{float}
\usepackage{enumitem}
\usepackage{forest}
\usepackage{algorithm}
\usepackage{algpseudocode}
\usepackage{cite}
\graphicspath{{img/}}

\sethlcolor{gray!20}

\def\bitcoin{%
  \leavevmode
  \vtop{\offinterlineskip 
    \setbox0=\hbox{B}%
    \setbox2=\hbox to\wd0{\hfil\hskip-.03em
    \vrule height .3ex width .15ex\hskip .08em
    \vrule height .3ex width .15ex\hfil}
    \vbox{\copy2\box0}\box2}}

\hypersetup{
colorlinks=false,
pdftitle={OP_RAND},
pdfpagemode=FullScreen,
}
\title{Emulating \texttt{OP\_RAND} in Bitcoin}

\author{Rarimo Protocol \\\\ Oleksandr Kurbatov \\ \small {ok@distributedlab.com}\\}

\date{}

\begin{document}
\maketitle

\begin{abstract}
This paper proposes a method of emulation of \verb|OP_RAND| opcode on Bitcoin through a trustless interactive game between transaction counterparties. The game result is probabilistic and doesn't allow any party to cheat, increasing their chance of winning on any protocol step. The protocol can be organized in a way unrecognizable to any external party and doesn't require some specific scripts or Bitcoin protocol updates. We will show how the protocol works on the simple \textbf{Thimbles Game} and provide some initial thoughts about approaches and applications that can use the mentioned approach. 
\end{abstract}

\section{Introduction}\label{sec:intro}
Bitcoin \cite{bitcoin} is a decentralized payment system that focuses on censorship resistance and cryptographic control of funds. Its payments have limited programmability, allowing the realization of the simplest spending conditions: signatures, hash, and time locks. Bitcoin script doesn't directly allow putting randomness and constructing the spending flow based on that. So, realizing the flow "\textit{Alice and Bob put for 5 BTC each, and Bob takes everything if the coin comes up tails}" wasn't possible upon the following assumptions:
\begin{enumerate}[noitemsep]
    \item The transaction can't derive or take randomness from somewhere at the moment of confirmation
    \item Bitcoin Script can't inspect the block, past or future transactions
    \item Each party can receive the same stack's state after each opcode processing
    \item We can't control the ECDSA or Schnorr signature determinism
    \item Bitcoin doesn't support \verb|OP_RAND| opcode =)
\end{enumerate}

All the limitations mentioned led to the situation where we couldn't find trustless solutions that allow scrambling randomness and use it for the protocol operating with bitcoins. This paper proposes a way to organize it via a 2-party interactive protocol and shows how these properties can be applied in the example of a thimbles game that takes bids in \bitcoin.

\subsection{Related work}
We list the set of protocols we inspired to construct the approach described in the paper. First of all, it's HTLC contracts and Lightning Network introduced by \cite{lightning}. These technologies allow you to lock bitcoins on the multisig address and then send payments off-chain, changing only a reconciliation transaction. We could initially use these properties to lock coins, allowing us to unlock them after the locktime if the game doesn't start.

Then, we reuse the properties provided by Atomic Swaps technology, which allows us to swap assets between different chains using secret knowledge and its commitment as a locking condition (totally trustless). The same technology can be modified using adaptor signatures introduced by \cite{adaptor} and taprootized untraceable version proposed by \cite{tas}. 

\section{Preliminaries}
$\mathbb{G}$ a cyclic group of prime order $p$ written additively, $G \in \mathbb{G}$ is the group generator. $a \in \mathbb{F}_p$ is a scalar value and $A \in \mathbb{G}$ is a group element. $\mathsf{hash}_p(m) \rightarrow h\in \mathbb{F}_p$ is the cryptographic hash function that takes as an input an arbitrary message $m$ and returns the field element $h$. $\mathsf{hash}_{160}(P) \rightarrow \mathsf{addr}\in \mathcal{A}$ is the function of hashing the public key with sha-256 and ripemd160 functions and receiving a valid bitcoin address as an output.

We define the relation for the proof $\pi$ as $\mathcal{R} = \{(w;x) \in \mathcal{W} \times \mathcal{X}: \phi_1(w,x), \phi_2(w,x) , \dots, \phi_m(w,x)\}$, where $w$ is a witness data, $x$ is a public data and $\phi_1(w,x), \phi_2(w,x) , \dots, \phi_m(w,x)$ the set of relations must be proven simultaneously.

We define a bitcoin transaction as $\mathsf{TX}\{(\mathsf{id, i, proof})^{(n)};(\mathsf{a \bitcoin, cond})^{(m)}\}$ with $n$ inputs and $m$ outputs, where $\mathsf{id}$ is the hash of the previous transaction, $i$ - output's index, $\mathsf{proof}$ - the list of data which is needed to transaction spending, $a$ - the number of coins in the output, $\mathsf{cond}$ - scriptPubKey conditions. For example, the P2PKH method requires $\mathsf{proof} \leftarrow \langle  \mathsf{PK}, \sigma\rangle$ and $\mathsf{cond}\leftarrow \langle $ \verb|OP_DUP, OP_HASH160,| $\mathsf{addr},$ \verb|OP_EQUALVERIFY, OP_CHECKSIG| $\rangle$. We are going to simplify the condition notation above to $\mathsf{addr}$ when referring to the P2PKH approach.

Let's note that the $\mathsf{proof}$ data isn't covered by the signature (we are referring to SegWit \cite{segwit}) because it's allocated in the witness data. So the signature $\sigma$ for $\mathsf{TX}\{(\mathsf{id, i, \textcolor{red}{proof}})^{(n)};(\mathsf{a \bitcoin, cond})^{(m)}\}$ and $\mathsf{TX}\{(\mathsf{id, i, \textcolor{blue}{-}})^{(n)};(\mathsf{a \bitcoin, cond})^{(m)}\}$ is equivalent.

\section{EC Point covenant}
First of all, let's see how we can implement the transaction with two counterparties and the following conditions: "It's possible to spend the second transaction output only in the case the first is spent". Traditionally, it could be organized using a hash lock contract, but 1 -- it's recognizable; 2 -- it won't help us to implement the final game.

\begin{algorithm}[H] \small
\caption{Creating the output that can be spent in the case of spending another output}
\label{alg:deposit}
\textbf{Condition:} Alice and Bob deposit 1\bitcoin \ each. Bob must be able to spend his 1\bitcoin \ only if Alice spends her 1\bitcoin. Bob's public key $P_b$ is known in advance.\\\\
\textbf{Flow:} 
\begin{algorithmic}
    \State \qquad 1. Alice generates:
    \begin{gather*}
        sk_a \xleftarrow{R} \mathbb{F}_p \\
        P_a = sk_aG \\
        \mathsf{addr}_a = \mathsf{hash}_{160}(P_a) \\
        C = \mathsf{hash}_p(P_a)\cdot G
    \end{gather*}
    \qquad and creates a proof $\pi_c$ for the relation:
    \begin{gather*}
        \mathcal{R}_{c} = \{P_a; \mathsf{addr}_a, C, G: \mathsf{hash}_{160}(P_a) \to \mathsf{addr}_a \ \land \ \mathsf{hash}_p(P_a)\cdot G \to C\}
     \end{gather*}
     \State \qquad 2. Bob verifies the proof $\pi_c$, takes $C$ and calculates:
     \begin{gather*}
         \mathsf{addr}_b = \mathsf{hash}_{160}(P_b + C)
     \end{gather*}
     \State \qquad 3. Bob creates a transaction and sends it to Alice:
     \begin{gather*}
         \mathsf{TX}_1\{(\mathsf{prev_A, i_A, \color{red}{-}}), (\mathsf{prev_B, i_B, \color{blue}{\sigma_B}(\mathsf{TX_1})});(\mathsf{1 \bitcoin, \color{red}{addr_a}}), (\mathsf{1 \bitcoin, \color{blue}{addr_b}})\}
     \end{gather*}
    \State \qquad 4. Alice co-signs the transaction and propagates it to the network:
    \begin{gather*}
         \mathsf{TX}_1\{(\mathsf{prev_A, i_A, \color{red}{\sigma_a}(\mathsf{TX_1})}), (\mathsf{prev_B, i_B, \color{blue}{\sigma_B}(\mathsf{TX_1})});(\mathsf{1 \bitcoin, \color{red}{addr_a}}), (\mathsf{1 \bitcoin, \color{blue}{addr_b}})\}
     \end{gather*}
\end{algorithmic}
\end{algorithm}

If Alice wants to spend her output, she needs to create a transaction and reveal a public key $P_a$ and the signature value.
\begin{gather*}
    \mathsf{TX}_2\{(\mathsf{TX}_1, 1, \color{red}{\langle P_a, \sigma_{P_a}(\mathsf{TX_2})\rangle});(\mathsf{1 \bitcoin, \color{red}{addr_{a'}}})\}
\end{gather*}

After the transaction is published, Bob can extract $P_a$ and recover the $\mathsf{hash}_p(P_a)$ value. Then the secret key for the second output is calculated as $sk = \mathsf{hash}_p(P_a) + sk_b$ (only Bob controls $sk_b$), and Bob can construct the signature related to $P_b+C$ public key and corresponding address.
\begin{gather*}
    \mathsf{TX}_3\{(\mathsf{TX_1, 2}, \color{blue}{\langle P_b + C, \sigma_{P_b + C}(\mathsf{TX_3})\rangle});(\mathsf{1 \bitcoin, \color{blue}{addr_{b'}}})\}
\end{gather*}

So, we have built the first part needed for emulating the randomness and our thimbles game. We need to note that in the previous example, if Alice doesn't spend her output and doesn't publish $P_a$ anywhere, Bob can't recover the key and spend his output as well. If we need to provide an ability to spend these outputs after some time (if the game hasn't started), we can do it through timelock conditions.
\begin{gather*}
    \mathsf{TX}_1\{(\mathsf{prev_A, i_A, \color{red}{\sigma_a}(\mathsf{TX_1})}), (\mathsf{prev_B, i_B, \color{blue}{\sigma_B}(\mathsf{TX_1})});(\mathsf{1 \bitcoin, \color{red}{addr_a} \lor \color{red}{addr_a'} + t_1}), (\mathsf{1 \bitcoin, \color{blue}{addr_b} \lor \color{blue}{addr_b'} + t_2})\}
\end{gather*}

\section{\texttt{OP\_RAND} emulation protocol}
We propose to emulate the \texttt{OP\_RAND} opcode with an interactive protocol between parties involved in the transaction. Introducing the Challenger $\mathcal{C}$ and Accepter $\mathcal{A}$ roles we can define the \texttt{OP\_RAND} emulation protocol as follows:
\begin{enumerate}
    \item $\mathcal{C}$ and $\mathcal{A}$ have their cryptographic keypairs $\langle sk_{\mathcal{C}}, P_{\mathcal{C}}\rangle$ and $\langle sk_{\mathcal{A}}, P_{\mathcal{A}}\rangle$. Only $P_{\mathcal{C}}$ value is public
    \item $\mathcal{C}$ generates the set of random values $a_1, a_2,\dots, a_n$ and creates a first rank commitments for them as $A_i = a_iG, i\in[1, n]$
    \item $\mathcal{C}$ selects one commitment $A_x$, assembles it with own public key as $R_{\mathcal{C}} = P_{\mathcal{C}}+A_x$ and publishes only the hash value of the result $\mathsf{hash}(R_{\mathcal{C}})$
    \item $\mathcal{C}$ creates second rank commitments as $h_i = \mathsf{hash}(A_i), i \in[1,n]$ and third rank commitments as $H_i = h_iG, i \in[1,n]$
    \item $\mathcal{C}$ creates a proof $\pi_a$ that all third rank commitments were derived correctly, and one of the first rank commitments is used for assembling with $P_{\mathcal{C}}$
    \item $\mathcal{C}$ proposes the set of third rank commitments to the $\mathcal{A}$ and provides $\pi_a$
    \item $\mathcal{A}$ verifies the proof $\pi_a$ and selects one of the third-rank commitments $H_y$ to assemble it with $P_{\mathcal{A}}$. The result $R_{\mathcal{A}}=P_{\mathcal{A}}+H_y$ is hashed $\mathsf{hash}(R_{\mathcal{A}})$ and published
    \item $\mathcal{A}$ creates a proof $\pi_r$ that one of the third rank commitments was used for assembling with $P_{\mathcal{A}}$ and sends it to $\mathcal{C}$. Additionally, the proof covers the knowledge of the discrete log of $P_{\mathcal{A}}$
    \item $\mathcal{C}$ verifies the proof $\pi_r$ and if it's valid publishes the $R_{\mathcal{C}}$
    \item $\mathcal{A}$ calculates $A_x = R_{\mathcal{C}}-P_{\mathcal{C}}$
    \item If $\mathsf{hash}(A_x)\cdot G = H_y$, $\mathcal{A}$ won. Otherwise lost 
\end{enumerate}

\section{An example of Thimbles Game}
Finally, we can show how the interactive protocol we introduced allows the organization of a trustless thimbles game between two counterparties. So, having Alice and Bob, the game could be described as follows:
\begin{enumerate}
    \item Alice generates two values and selects one of them (don't reveal the selected value to Bob). In other words, Alice chooses a thimble with a ball under it
    \item Alice locks her coins with TX in a way that can be unlocked by publishing the selected value
    \item Bob selects the thimble: takes one value from the proposed by Alice (and also doesn't reveal it). Then, Bob constructs the address using her public key and selected value
    \item Bob creates a TX that requires Alice's input and pays to Bob's new address or Alice after locktime 
    \item Alice reveals the value she selected initially by co-signing and publishing the TX created by Bob
    \item If Bob selected the same value --- he can take all coins. If not, Alice can spend coins after locktime
\end{enumerate}

\begin{algorithm}[H] \small
\caption{Thimbles game}
\label{alg:deposit}
\textbf{Condition:} Alice and Bob deposit 5\bitcoin \ each ($\langle\mathsf{prev_a, i_a}\rangle, \langle\mathsf{prev_b, i_b}\rangle$ are appropriate unspent outputs). Bob can take all coins only in the case he guesses the value selected by Alice. \\\\
\textbf{Flow:} 
\begin{algorithmic}
    \State \qquad 1. Alice generates:
    \begin{gather*}
        sk_a \xleftarrow{R} \mathbb{F}_p \\
        P_a = sk_aG \\
        a_1, a_2 \xleftarrow{R} \mathbb{F}_p \\
        A_1 = a_1G, A_2 = a_2G \\
        h_1 = \mathsf{hash}_p(A_1), h_2 = \mathsf{hash}_p(A_2) \\
        H_1 = h_1G, H_2 = h_2G \\
        \mathsf{addr}_a = \mathsf{hash}_{160}(P_a + A_1)
    \end{gather*}
    \qquad and creates a proof $\pi_a$ for the relation:
    \begin{gather*}
        \mathcal{R}_{a} = \{a_1, a_2; H_1, H_2, P_a, G, \mathsf{addr}_a: 
        \\ a_1G \to A_1 \ \land \ a_2G \to A_2 \ \land \ \mathsf{hash}_p(A_1) \to h_1 \ \land \ \mathsf{hash}_p(A_2) \to h_2 \ \land
        \\ h_1G \to H_1 \ \land \  h_2G \to H_2 \ \land
        \\ (\mathsf{hash}_{160}(P_a + {\color{red}{A_1}}) \to \mathsf{addr}_a \ \lor \  \mathsf{hash}_{160}(P_a + A_2) \to \mathsf{addr}_a)
        \}
    \end{gather*}
    \State \qquad 2. Alice creates a transaction
    \begin{gather*}
        \mathsf{TX}_1\{(\mathsf{prev_A, i_A, \color{red}{-}});(\mathsf{5 \bitcoin, \color{red}{addr_a}})\}
    \end{gather*}
    \State \qquad 3. Bob generates $sk_b \xleftarrow{R} \mathbb{F}_p, P_b = sk_bG$, verifies the proof $\pi_a$, takes $H_1, H_2$ and selects only one value of them ($H_1$ for example). Then Bob generates an address as:
    \begin{gather*}
        \mathsf{addr}_b = \mathsf{hash}_{160}(P_b + H_1)
    \end{gather*}
    calculates the signature proving the knowledge of $sk_b$:
    \begin{gather*}
        \mathsf{\sigma} \leftarrow \mathsf{sigGen}(sk_b, \mathsf{addr}_b)
    \end{gather*}
    and generates a proof $\pi_r$ for the relation:
    \begin{gather*}
        \mathcal{R}_{r} = \{P_b, \sigma; \mathsf{addr}_b, H_1, H_2: \\\mathsf{sigVer}(\sigma, P_b, \mathsf{addr}_b) \to \mathsf{true} \ \land (\mathsf{hash}_{160}(P_b + H_1) \to \mathsf{addr}_b \ \lor \ \mathsf{hash}_{160}(P_b + H_2) \to \mathsf{addr}_b)\}
     \end{gather*}
     \State \qquad 4. Bob creates the TX in the following way:
     \begin{gather*}
        \mathsf{TX}_2\{(\mathsf{TX_1, 1, \color{red}{-}}), (\mathsf{prev_B, i_B, \color{blue}{\sigma}_B(TX_2)});(\mathsf{10 \bitcoin, \color{blue}{addr_b}} \lor \textcolor{red}{P_a + t_1})\}
    \end{gather*}
    \State \qquad 5. Alice verifies the proof $\pi_r$ and complete the transaction with:
    \begin{gather*}
        \mathsf{TX}_2\{(\mathsf{TX_1, 1, \langle \color{red}{P_a + A_1, \sigma_{P_a + A1}(TX_2)}\rangle}), (\mathsf{prev_B, i_B, \color{blue}{\sigma}_B(TX_2)});(\mathsf{10 \bitcoin, \color{blue}{addr_b}} \lor \textcolor{red}{P_a + t_1})\}
    \end{gather*}
    \State \qquad 6. Alice propagates both $\mathsf{TX_1}$ and $\mathsf{TX_2}$ to the network.
\end{algorithmic}
\end{algorithm}

So Bob doesn't know which value was selected by Alice, Alice doesn't know what Bob selected. When Alice spends the output from the $\mathsf{TX_1}$, she publishes $P_a + A_1$ value. Bob knows $P_a$ so he can easily recover $A_1$ value and corresponding $h_1 = \mathsf{hash}_p(A_1)$. 

If the secret key $h_1 + sk_b$ satisfies the address $\mathsf{addr}_b$, Bob can take 10\bitcoin, from the $\mathsf{TX_2}$. If not --- Alice can take them after the timelock.

\begin{figure}[H]
    \centering
    \includegraphics[width=1\linewidth]{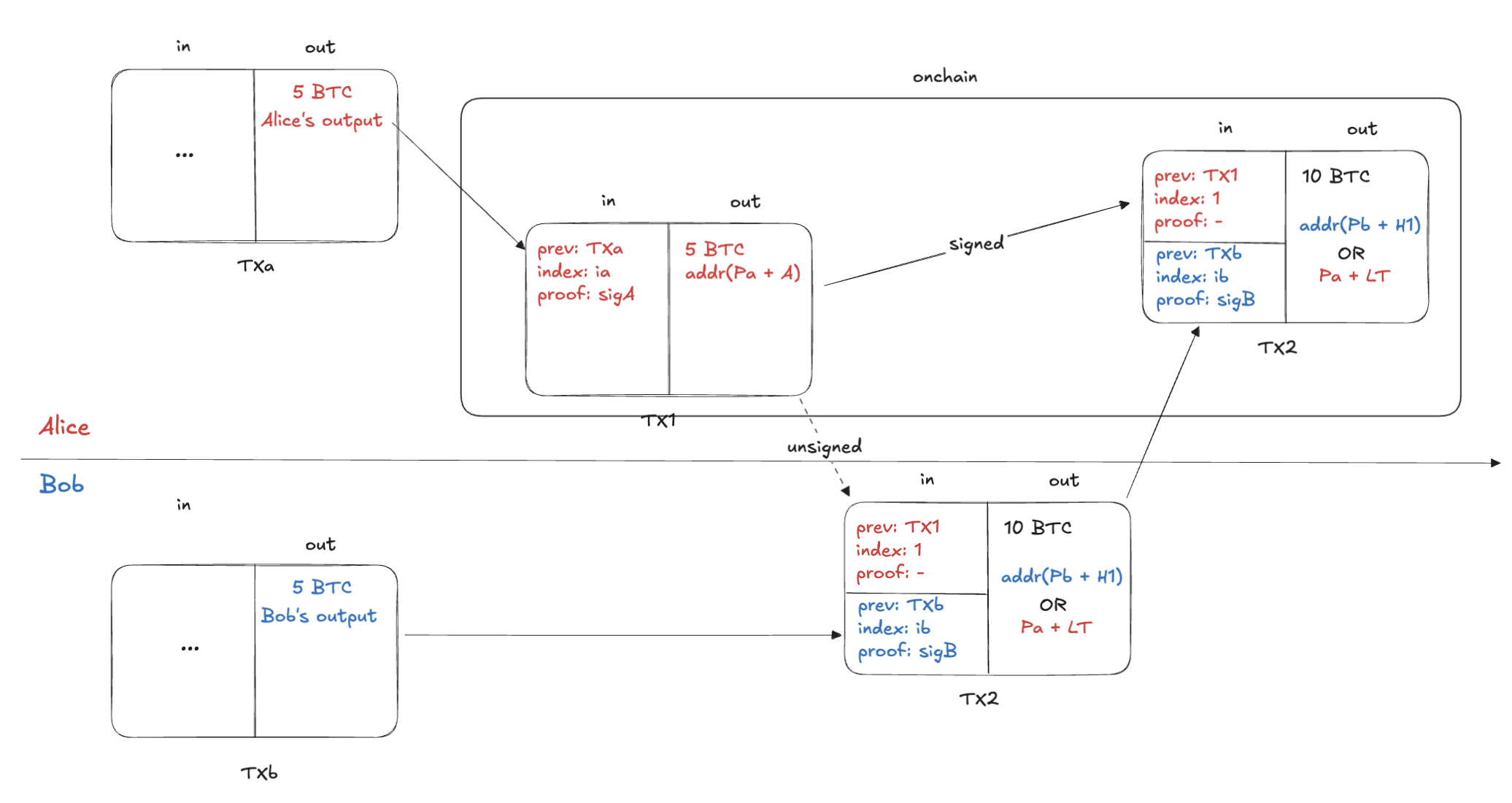}
    \caption{Transactions flow}
    \label{fig:full-flow}
\end{figure}

\section{Future work}
Although it is quite difficult to find cases that require randomness in Bitcoin, we found that the parts described in the paper can be used to constrain the potential states and their sequence. In other words, we can emulate a minimalistic virtual machine within transaction UTXOs.

Imagine Alice has some hidden state $s$ she doesn't want to reveal in advance. There is a set of operations $f_1(), f_2()$ (we reduce the number of possible operations to 2 for simplicity), which can be applied to the state. So, we have a limited set of states we can receive after the particular function is executed.
\begin{figure}[H]
    \centering
    \includegraphics[width=0.8\linewidth]{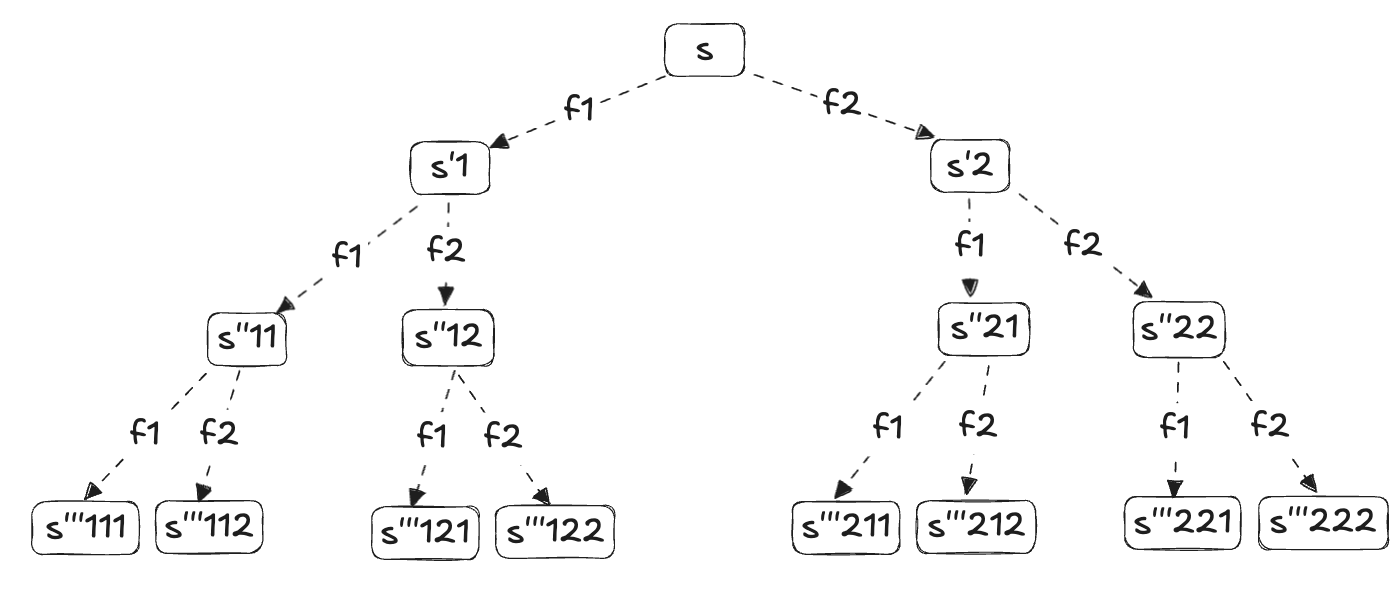}
    \caption{Alternative execution traces}
    \label{fig:state-tree}
\end{figure}
Then Alice does the following:
\begin{enumerate}
    \item Creates a commitment $P_0 = P_a + sG$
    \item Creates the set of commitments for possible transitions: 
\begin{gather*}
    P_1 = P_a+\mathsf{hash}(P_0)G+f_1(s)G\\P_2 = P_a+\mathsf{hash}(P_0)G+f_2(s)G
\end{gather*}
    \item Commitment for the following layer:
\begin{gather*}
    P_{11} = P_a+\mathsf{hash}(P_1)G+f_1(f_1(s))G\\P_{12} = P_a+\mathsf{hash}(P_1)G+f_1(f_2(s))G\\
    P_{21} = P_a+\mathsf{hash}(P_2)G+f_2(f_1(s))G\\P_{22} = P_a+\mathsf{hash}(P_1)G+f_2(f_2(s))G
\end{gather*}
    \item And finally, for the last layer:
\begin{gather*}
    P_{111} = P_a+\mathsf{hash}(P_{11})G+f_1(f_1(f_1(s)))G\\
    ... \\
    P_{222} = P_a+\mathsf{hash}(P_{22})G+f_2(f_2(f_2(s)))G
\end{gather*}
    \item Then Alice generates the $\mathsf{addr}$ for each commitment and the proof that all commitments and addresses were generated correctly.
    \item Then Alice creates the transaction:
    \begin{align*}
        \mathsf{TX}\{(\mathsf{prev_A, i_A, \color{black}{\sigma_a(TX)}});&(\mathsf{1 \bitcoin, \color{black}{addr_0} + \mathsf{hashlock}(s)}),
        \\ &(\mathsf{1 \bitcoin, \color{black}{addr_1} \ OR \ \color{black}{addr_2}})
        \\ &(\mathsf{1 \bitcoin, \color{black}{addr_{11}} \ OR \ \color{black}{addr_{12}} \ OR \ \color{black}{addr_{21}} \ OR \ \color{black}{addr_{22}}})
        \\ &... \}
    \end{align*}
    \item Note that all addresses can be put as alternative spending paths in the Taproot. After the transaction is submitted, Alice can spend outputs one by one, proving the correctness of all states regardless of the exact trace that was executed.
\end{enumerate}
\begin{figure}[H]
    \centering
    \includegraphics[width=0.8\linewidth]{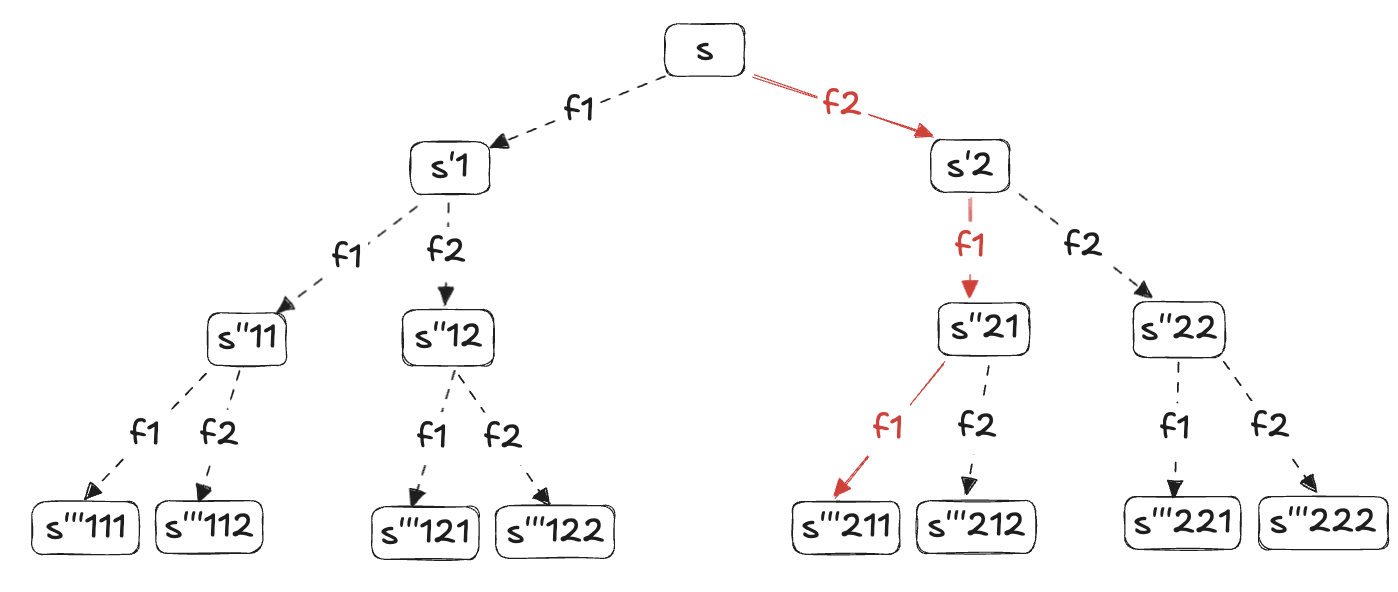}
    \caption{Actual state transitions}
    \label{fig:execution-trace}
\end{figure}
We see potential in the mentioned approach, but still, a huge number of improvements and potential risks MUST be considered. 

\section*{Acknowledgments}
Special thanks to Tadge Dryja for throwing challenging cryptographic tasks and Anthony Towns for the solution's safety.

\end{document}